\documentclass[%
 aip,
 jmp,%
 amsmath,amssymb,
 reprint,%
]{revtex4-2}

\usepackage{graphicx}% Include figure files
\usepackage{dcolumn}% Align table columns on decimal point
\usepackage{bm}% bold math
\usepackage{mathptmx} % times new roman fonts
\usepackage{amsmath,amssymb}
\begin{document}

\title[Plasmonic Quantum Yield Enhancement of a Single Molecule Near a Nanoegg]
{Plasmonic Quantum Yield Enhancement of a Single Molecule Near a Nanoegg} 

\author{Luke C. Ugwuoke}
\affiliation{Department of Physics, University of Pretoria.\\
Private bag X20, Hatfield 0028, South Africa}
\email{luke.ugwuoke@up.ac.za}
\author{Tom\'{a}\v{s} Man\v{c}al}
\affiliation{Faculty of Mathematics and Physics, Charles University.\\
Ka Karlovu 5, 121 16 Prague 2, Czech Republic}
\author{Tjaart P. J. Kr\"{u}ger}
\affiliation{Department of Physics, University of Pretoria.\\
Private bag X20, Hatfield 0028, South Africa}
\email{luke.ugwuoke@up.ac.za}

\date{\today}

\begin{abstract}
We investigate the impact of the dipole-active modes formed via the mode-mixing of the dipole mode with higher-order 
surface plasmon modes of a nanoegg on the radiative decay rate and quantum yield of an excited molecule near the nanoegg. 
The Purcell factor, rate of power dissipation by the emitter, antenna efficiency of the nanoegg, as well as quantum yield enhancement of the emitter, were studied using the quasistatic approximation and the semiclassical theory of radiation, following the Gersten-Nitzan and Ford-Weber approaches. 
Compared to the concentric nanoshell, we show that the dielectric core-metallic shell nanoegg is a more efficient plasmonic nanoantenna for radiative decay rate enhancement of single emitters. The quantum yield of the emitter was found to be more enhanced near the nanoshell, while its emission rate was found to be more enhanced near the nanoegg.  
\end{abstract}

                             
\maketitle

\section{Introduction}\label{s1}
Theoretical models describing the modifications of the decay rate of an excited molecule near a metal layer can be traced back
to Chance et al. \cite{CPS78}. Similar models were then proposed by Gersten and Nitzan \cite{GN81}, Ruppin \cite{Rupp82}, and 
Ford and Weber \cite{FW84}, for metallic nanospheroids and nanospheres respectively. These models predict that the decay rate 
of an excited molecule near a metal-dielectric interface increases with decreasing metal-molecule separation and oscillates 
with increasing metal-molecule distance \cite{CPS78,GN81,Rupp82,FW84}. The former is due to the increased rate of non-radiative energy 
transfer from the excited molecule to the metal, while the latter is due to interference between the incident field on the 
molecule and the induced field on the molecule as a result of the reflected field at the metal-dielectric boundary. 

The enhancement of the radiative decay rate of a molecule near a metal is known as the Purcell effect 
\cite{Ged02,Anger06,BN07}. It has been studied both theoretically and experimentally in the emission stage of the phenomenon of 
plasmon-enhanced fluorescence (PEF). PEF, the effect of increase in the emission rate of the molecule, is characterized by an increase 
in both the quantum yield \cite{Mert07,Mert09,Wien14} and excitation rate of the molecule \cite{Anger06,Khat14,Chris16}. PEF reaches its
maximum at a wavelength red-shifted from the dipolar localized surface plasmon resonance (LSPR) of the metal nanoparticle (MNP) \cite{BN07,Thom04}.
 It depends on the excitation wavelength, the optical properties of the molecule, the molecule's dipole orientation, the molecule's position from the MNP, the MNP-molecule distance, the MNP geometry, the material composition of the MNP, polarization of the incident electric field, and the dielectric embedding medium. However, molecule-dependent, plasmon-induced quenching of fluorescence can also occur at short distances between the MNP surface and the molecule\cite{Anger06}.

The dependence of PEF on MNP geometry has been investigated in molecules near metal layers \cite{CPS78,Ged02,Ender19,Mack08}, 
metallic spheres \cite{Anger06,BN07,Mert07,Fu07,Sun09,Car10,Zhang10,Beyer11,Guz12,Pun13,Maj17}, 
metallic nanorods and nanospheroids \cite{Wien14,Khat14,Moha08,Mert09,Bard09,Ming12,Buja14,Wang16,Faro19}, 
metallic nanoshells \cite{Chris16,Bard09,Bard08,Fu12}, nanoparticle dimers \cite{Musk07,Liaw09,Kami17,Xin19}, 
and with MNPs of irregular geometries \cite{Thom04,Kern12}. The molecule-dependence of PEF has been studied with different 
molecules, including photosynthetic pigments \cite{Wien14,Mack08,Car10,Beyer11,Buja14,Faro19,Yang16},
where PEF has been shown to be more pronounced in weakly-emitting molecules \cite{Wien14,Khat14,Mert07}, 
while the material-dependence of PEF has been mostly reported using gold \cite{Anger06,Bard08,Pun13,Faro19}, silver \cite{Chris16,Fu07,Guz12}, and aluminum \cite{Chow09} nanoparticles. The modification of the radiative decay rate of fluorophores near MNPs has several promising applications such as in the design of optical devices for fluorescence microscopy \cite{Wei17}, photocurrent enhancement in biosolar cells \cite{Yang16,Stand09,Wu11}, and increased signal detection in biosensors \cite{Bau13}.

Nanoeggs belong to a group of tunable asymmetric nanostructures capable of supporting multiple LSPR \cite{Wang06,Wu06,Night08,ZaZa13,Nort16,LTT19}. Their plasmonic behaviour is attributed to the plasmon hybridization of 
solid and cavity plasmons with different angular momentum numbers, which is symmetry-forbidden in concentric nanoshells \cite{Wang06,Wu06}.
A similar description suggests that off-setting the core of a concentric nanoshell causes the dipolar surface plasmon mode to couple to higher-order multipoles \cite{ZaZa13,Nort16}, leading to the formation of dipole-active modes. The excitation stage of PEF, which is characterized by local field enhancements, has been investigated for nanoeggs \cite{Wu06,Nort16}, and it was found that their field enhancement factors increase with increasing core-offset. On the other hand, single-particle spectroscopy studies have shown that the LSPR of nanoeggs undergoes a redshift as the core-offset increases \cite{Wang06,Night08,ZaZa13,Nort16}. 

The emission stage of PEF, which is characterized by enhancement of the intrinsic quantum yield of a molecule, has not been reported previously, for an excited molecule near a nanoegg. 
The present study focuses on the emission stage of an excited molecule near a dielectric core-metallic shell (DCMS) nanoegg, surrounded by a
dielectric medium. The aim of this study is to investigate the impact of the dipole-active modes on the radiative decay rate and quantum yield of the molecule.

\begin{figure}[h!]
	\centering
	\includegraphics[width=0.7\textwidth]{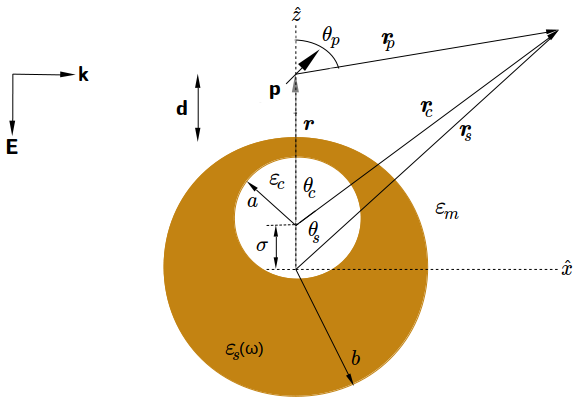}
	\caption{\small{Model geometry of the molecule-nanoegg system. The system is surrounded by a dielectric medium of dielectric 
			constant $\varepsilon_{m}$. The nanoegg consists of a metallic shell of dielectric constant $\varepsilon_{s}(\omega)$, 
			an off-centre core of dielectric constant $\varepsilon_{c}$, and a core-offset $\sigma$. The molecule, modelled as an electric point dipole with a dipole moment $\mathbf{p}$, is at a distance $\mathbf{d} = d\hat{z}$ from the shell surface, and oriented at a polar angle $\theta_{p}$ from the $z$ axis. 
	}}
	\label{f1}
\end{figure}

\section{Theory}\label{s2}
We will use the theoretical method proposed by Gersten and Nitzan \cite{GN81} and Ford and Weber \cite{FW84}. 
Their method is based on the electrostatic approximation and the semiclassical theory of radiation. 
It involves the following:
the excited molecule is treated as an oscillating point dipole which provides the source field, both the MNP-molecule distance and the MNP size have to be small compared to the wavelength of light in the medium, the local, wavevector-independent, complex dielectric function of the metal is used, the electric potentials are solutions of Laplace equation in each region of interest, the decay rates are obtained from the modified classical power of the molecular dipole via the correspondence principle, and only the dipolar surface plasmon mode contributes to Purcell effect. 

In addition to the aforementioned method, the solid-harmonic addition theorem (SHAT) \cite{Nort16,Cala78} will be used to express the shell 
coordinates in terms of the core coordinates at the core-shell interface of the nanoegg. We will also assume that the molecule is positioned near the surface of the nanoegg in the direction of the core-offset as shown in Fig. \ref{f1}. Two different orientations of 
the molecular dipole with respect to the nanoegg surface will be considered. 
The perpendicular orientation, which constitutes the maximum contribution to 
enhancement factors, is often compared to experimental values \cite{Wien14,Khat14}, and the parallel orientation, which gives the minimum 
contribution \cite{GN81,Mert09,Moroz11}.

We will consider a nanoegg with a gold shell and a silica core. 
In the local response approximation (LRA), a Drude-Lorentz dielectric function for gold, which agrees with experimental data in the 
wavelength region $500$ nm to $1000$ nm, has been proposed as follows\cite{Vial05}:
\begin{equation}\label{e1}
\varepsilon_{s}(\omega) = \varepsilon_{\infty}-\frac{\omega^{2}_f}{\omega(\omega+i\gamma_{f})}-
\frac{f\omega^{2}_b}{\omega(\omega+i\gamma_{b})-\omega^{2}_{b}},
\end{equation}
with $\varepsilon_{\infty} = 5.9673, \omega_{f} = 8.7411$ eV, $\gamma_{f} = 0.0658$ eV, $\omega_{b} = 2.6885$ eV, $\gamma_{b} = 0.4337$ eV, 
and $f = 1.09$. $\varepsilon_{\infty}$ is the high-frequency dielectric constant of gold, which accounts for the polarization of the positive ion core, $\omega_{f}$ and $\gamma_{f}$ are the plasma frequency and damping rate of the free electrons respectively, 
$\omega_{b}$ and $\gamma_{b}$ are the resonance frequency and damping rate of the bound electrons respectively, $f$ is the oscillator
strength, and $\omega$ is the frequency of the incident light. 

In the calculation of Purcell factors, we will correct the quasistatic dipole polarizability of the nanoegg for radiation damping. 
Radiation damping is due to a radiation reaction field produced by the induced dipole moment on the MNP \cite{Anger06,Moroz11}. 
Some authors have prescribed a method that takes into account the first-order correction to the quasi-static polarizability due to 
radiation reaction. Without this correction, the optical theorem is violated \cite{Anger06,Mert07,ZaZa13}. However, in the calculation of the 
non-radiative energy transfer rate, we shall only correct the dipole term of the quasistatic multipole polarizability of the nanoegg 
for radiation damping. This approach is appropriate for MNPs less than $80$ nm \cite{Mert07,Moroz11}. 

\subsection{Perpendicular Dipole}\label{s2.1}
When the molecular dipole is normal to the surface of the MNP, both the dipole potential and the electric potentials in the core, shell,
and medium regions of the MNP, are independent of the azimuth angle $\phi$ of the dipole \cite{FW84,Maj17}. For the 
molecule-nanoegg system, these potentials can therefore be written as \cite{FW84,Nort16}
\begin{eqnarray}
\Phi_{c}(r_{c},\theta_{c}) = \sum_{n=1}^{\infty}A_{n}\Big(\frac{r_{c}}{a}\Big)^{n}P_{n}(\cos\theta_{c}),\label{e2}\\
\Phi_{s}(r_{s},\theta_{s}) = \sum_{n=1}^{\infty}\Big[B_{n}\Big(\frac{r_{s}}{b}\Big)^{n}+
C_{n}\Big(\frac{b}{r_{s}}\Big)^{n+1}\Big]P_{n}(\cos\theta_{s}), \label{e3}\\
\Phi_{m}(r_{s},\theta_{s}) = \Phi_{dip}(r_{p},\theta_{p})+\Phi_{ind}(r_{s},\theta_{s}), \label{e4}
\end{eqnarray}
where \cite{FW84,Maj17}
\begin{equation}\label{e5}
\Phi_{dip}(r_{p},\theta_{p}) = \frac{\mathbf{p}.\mathbf{z}}{\varepsilon_{m}r^{3}_{p}} = 
\frac{\text{p}_{z}\cos\theta_{p}}{\varepsilon_{m}r^{2}_{p}} = 
\sum_{n=1}^{\infty}E_{n}\Big(\frac{r_{s}}{b}\Big)^{n}P_{n}(\cos\theta_{s}),
\end{equation}
with 
\begin{equation}\label{e6}
E_{n} = -\frac{\text{p}_{z}(n+1)b^{n}}{\varepsilon_{m} r^{n+2}}, r = b+d,
\end{equation}
and 
\begin{equation}\label{e7}
\Phi_{ind}(r_{s},\theta_{s}) = \sum_{n=1}^{\infty}D_{n}\Big(\frac{b}{r_{s}}\Big)^{n+1}P_{n}(\cos\theta_{s}).
\end{equation}
Here, $r_{c}$ and $r_{s}$ have been normalized with their respective values at the boundaries, and $P_{n}(u)$ is the Legendre function 
of the first kind, obtained for $m=0$, which corresponds to the perpendicular dipole orientation.  $m$ is the azimuthal number. 
$A_{n}, B_{n}$ and $C_{n}$, and $D_{n}$ are the complex amplitudes of the electrostatic potential in the core, shell, medium regions 
of the nanoegg-emitter system respectively, for the normal dipole. $E_{n}$ is the amplitude of the source normal dipole potential. 

At the boundaries, both the potential and the normal component of the displacement field must be continuous, leading to
the following boundary conditions \cite{Nort16}: 
\begin{eqnarray}
\Phi_{c}\Big(r_{c},\theta_{c}\Big)\Big|_{r_{c}=a}
= \Phi_{s}\Big(r_{s},\theta_{s}\Big)\Big|_{r_{c}=a}, \label{e8}\\
\Phi_{s}\Big(r_{s},\theta_{s}\Big)\Big|_{r_{s} = b} = \Phi_{m}\Big(r_{s},\theta_{s}\Big)\Big|_{r_{s} = b}, \label{e9}\\
\varepsilon_{c}\dfrac{\partial \Phi_{c}\Big(r_{c},\theta_{c}\Big)}{\partial r_{c}}\Big|_{r_{c} = a} = 
\varepsilon_{s}(\omega)\dfrac{\partial \Phi_{s}\Big(r_{s},\theta_{s}\Big)}{\partial r_{c}}\Big|_{r_{c} = a},\label{e10}\\
\varepsilon_{s}(\omega)\dfrac{\partial \Phi_{s}\Big(r_{s},\theta_{s}\Big)}{\partial r_{s}}\Big|_{r_{s} = b} = 
\varepsilon_{m}\dfrac{\partial \Phi_{m}\Big(r_{s},\theta_{s}\Big)}{\partial r_{s}}\Big|_{r_{s} = b}. \label{e11}
\end{eqnarray}

Setting $u_{c} \equiv \cos\theta_{c}$ and $u_{s} \equiv \cos\theta_{s}$, and 
combining Eqs. (\ref{e2}-\ref{e7}) and Eqs. (\ref{e8}-\ref{e11}), we obtain:
\begin{eqnarray}
\sum_{n=1}^{\infty}A_{n}P_{n}(u_{c}) = \sum_{n=1}^{\infty}\Big[B_{n}\Big(\frac{r_{s}}{b}\Big)^{n}+
C_{n}\Big(\frac{b}{r_{s}}\Big)^{(n+1)}\Big]P_{n}(u_{s})\Big|_{r_{c} = a}, \label{e12}\\
\sum_{n=1}^{\infty}[B_{n}+C_{n}]P_{n}(u_{s}) = \sum_{n=1}^{\infty}[E_{n}+D_{n}]P_{n}(u_{s}), \label{e13}\\
\varepsilon_{c}\sum_{n=1}^{\infty}nA_{n}P_{n}(u_{c}) = 
a\varepsilon_{s}(\omega)\sum_{n=1}^{\infty} \dfrac{\partial}{\partial r_{c}} \Big[B_{n}\Big(\frac{r_{s}}{b}\Big)^{n}+
C_{n}\Big(\frac{b}{r_{s}}\Big)^{(n+1)}\Big]P_{n}(u_{s})\Big|_{r_{c} = a}, \label{e14}\\
\varepsilon_{s}(\omega)\sum_{n=1}^{\infty}[nB_{n}-(n+1)C_{n}]P_{n}(u_{s}) = 
\varepsilon_{m}\sum_{n=1}^{\infty}[nE_{n}-(n+1)D_{n}]P_{n}(u_{s}). \label{e15}
\end{eqnarray}

Multiplying both sides of Eqs. (\ref{e12} \& \ref{e14}) each by $P_{l}(u_{c})$
and  Eqs. (\ref{e13} \& \ref{e15}) each by $P_{l}(u_{s})$, and integrating each one respectively via 
\begin{equation}\label{e16}
\beta_{l} \int_{-1}^{1}P_{l}(u)P_{n}(u)du = \delta_{ln}, ~~\beta_{l} = l+\frac{1}{2},
\end{equation}
we obtain 
\begin{eqnarray}
A_{l} = \sum_{n=1}^{\infty}K_{ln}B_{n}+\sum_{n=1}^{\infty}M_{ln}C_{n}, \label{e17}\\
B_{l}+C_{l} = E_{l}+D_{l}, \label{e18}\\
\varepsilon_{c}lA_{l} = \varepsilon_{s}(\omega)\Big[\sum_{n=1}^{\infty}L_{ln}B_{n}+\sum_{n=1}^{\infty}N_{ln}C_{n}
\Big], \label{e19}\\
\varepsilon_{s}(\omega)[lB_{l}-(l+1)C_{l}] = 
\varepsilon_{m}[lE_{l}-(l+1)D_{l}], \label{e20}
\end{eqnarray}
where 
\begin{eqnarray}
K_{ln} = \frac{\beta_{l}}{b^{n}}\int_{-1}^{1}r_{s}^{n}P_{n}(u_{s})\Big|_{r_{c} = a}P_{l}(u_{c})du_{c}, \label{e21}\\
M_{ln} = \beta_{l}b^{n+1}\int_{-1}^{1}\frac{P_{n}(u_{s})}{r_{s}^{n+1}}\Big|_{r_{c}=a}P_{l}(u_{c})du_{c}, \label{e22}\\
L_{ln} = \frac{\beta_{l}a}{b^{n}}\int_{-1}^{1}
\dfrac{\partial }{\partial r_{c}}\Big[r_{s}^{n}P_{n}(u_{s})\Big]_{r_{c} = a}P_{l}(u_{c})du_{c}, \label{e23}\\
N_{ln} = \beta_{l}ab^{n+1}\int_{-1}^{1}
\dfrac{\partial }{\partial r_{c}}\left[\frac{P_{n}(u_{s})}{r_{s}^{n+1}}\right]_{r_{c}=a}P_{l}(u_{c})du_{c}. \label{e24}
\end{eqnarray}
In order to evaluate Eqs. (\ref{e21}-\ref{e24}), we need to make use of the SHAT 
in spherical coordinates. This theorem allows us to express the integrands in Eqs. (\ref{e21}-\ref{e24}) in terms of the core coordinates $(r_{c},u_{c})$ and the core-offset $\sigma$. 
The SHAT theorem states that given two off-centre spherical harmonic coordinates $\mathbf{r}$ and $\mathbf{r'}$, 
then for $m = 0$\cite{Cala78}
\begin{eqnarray}
R_{n}(\mathbf{r}+\mathbf{r'}) = \sum_{k = 0}^{n}
\left(\begin{array}{c}
n\\k
\end{array}\right)R_{k}(\mathbf{r})R_{n-k}(\mathbf{r'}), \label{e25}\\
S_{n}(\mathbf{r}+\mathbf{r'}) = \sum_{k = n}^{\infty}(-1)^{k-n} \label{e26}
\left(\begin{array}{c}
k\\n
\end{array}\right)S_{k}(\mathbf{r})R_{k-n}(\mathbf{r'}), 
\end{eqnarray}
where $R_{n}(\mathbf{r})$ and $S_{n}(\mathbf{r})$ are the interior and exterior solutions of the Laplace equation in spherical
coordinates, given as \cite{FW84,Cala78}
\begin{eqnarray}
R_{n}(\mathbf{r}) = r^{n}P_{n}(u), \label{e27}\\
S_{n}(\mathbf{r}) = \frac{1}{r^{n+1}}P_{n}(u), \label{e28}
\end{eqnarray}
where $\mathbf{r} = (r,u)$ and $u = \cos\theta$. 
From Fig. \ref{f1}, we have
$\mathbf{r}_{s} = \mathbf{R}+\mathbf{r}_{c}$, where $\mathbf{r}_{s} = (r_{s},u_{s}), 
\mathbf{r}_{c} = (r_{c},u_{c})$, and $\mathbf{R} = (\sigma,1)$, since the core-offset lies along the vertical. 
Then, setting $\mathbf{r} = \mathbf{r_{c}}$,
and $\mathbf{r'} = \mathbf{R}$ in Eqs. (\ref{e25} \& \ref{e26}), respectively,
and making use of Eqs. (\ref{e27} \& \ref{e28}) and $P_{n}(1)=1$, we obtain 
\begin{eqnarray}
r_{s}^{n}P_{n}(u_{s}) = \sum_{k = 0}^{n}
\left(\begin{array}{c}
n\\k
\end{array}\right)r_{c}^{k}P_{k}(u_{c})\sigma^{n-k}, \label{e29}\\
\frac{P_{n}(u_{s})}{r_{s}^{n+1}} = \sum_{k = n}^{\infty}(-1)^{k-n}
\left(\begin{array}{c}
k\\n
\end{array}\right)\frac{1}{r_{c}^{k+1}}P_{k}(u_{c})\sigma^{k-n}. \label{e30}
\end{eqnarray}
Substituting Eq. (\ref{e29}) in Eqs. (\ref{e21} \& \ref{e23}) respectively, and 
substituting Eq. (\ref{e30}) in Eqs. (\ref{e22} \& \ref{e24}) respectively, leads to 
\begin{eqnarray}
L_{ln} = lK_{ln},~~K_{ln} = \left(\begin{array}{c}
n\\l
\end{array}\right)\frac{a^{l}\sigma^{n-l}}{b^{n}}
\begin{cases}
1,& n \ge l\\
0,& n<l
\end{cases}, \label{e31}\\
N_{ln} = -(l+1)M_{ln},~~M_{ln} = (-1)^{l-n}\left(\begin{array}{c}
l\\n
\end{array}\right)\frac{b^{n+1}\sigma^{l-n}}{a^{l+1}}
\begin{cases}
1,& l \ge n\\
0,& l<n
\end{cases}. \label{e32}
\end{eqnarray}
Substituting Eq. (\ref{e31}) into Eq. (\ref{e19}), and Eq. (\ref{e32}) into Eq.(\ref{e19}), 
Eqs. (\ref{e17}-\ref{e20}) can be re-written as: 
\begin{eqnarray}
A_{l} = \sum_{n=1}^{N}K_{ln}B_{n}+\sum_{n=1}^{N}M_{ln}C_{n}, \label{e33}\\
B_{l}+C_{l} = E_{l}+D_{l}, \label{e34}\\
\varepsilon_{c}lA_{l} = \varepsilon_{s}(\omega)\Big[l\sum_{n=1}^{N}
K_{ln}B_{n}-(l+1)\sum_{n=1}^{N}M_{ln}C_{n}
\Big], \label{e35}\\
\varepsilon_{s}(\omega)[lB_{l}-(l+1)C_{l}] = 
\varepsilon_{m}[lE_{l}-(l+1)D_{l}], \label{e36}
\end{eqnarray}
where we have truncated the summation to some finite number $N$. 

To obtain the static multipole polarizability of the nanoegg, we need to express the amplitude of the induced potential $D_{l}$ 
in terms of the amplitude of the incident potential $E_{l}$. 
Eliminating $B_{l}$ using Eq. (\ref{e34}) and Eq. (\ref{e36}), we obtain 
\begin{equation}\label{e37}
D_{l} = \frac{ (2l+1)\varepsilon_{s}(\omega)C_{l}-E_{l}l\Big[\varepsilon_{s}(\omega)-\varepsilon_{m}\Big] }
{ l\varepsilon_{s}(\omega)+(l+1)\varepsilon_{m} }.
\end{equation}
Next, we eliminate
$A_{l}$ using Eq. (\ref{e33}) and Eq. (\ref{e35}), to obtain
\begin{equation}\label{e38}
0 = \sum_{n=1}^{N}K_{ln}B_{n}+\frac{
	\Big[l\varepsilon_{c}+(l+1)\varepsilon_{s}(\omega)\Big]}{l\Big[\varepsilon_{c}-\varepsilon_{s}(\omega)\Big]}\sum_{n=1}^{N}M_{ln}C_{n}. 
\end{equation}
Then we eliminate $D_{l}$ using Eq. (\ref{e34}) and Eq. (\ref{e36}), to find 
\begin{equation}\label{e39}
B_{l} = \frac{(2l+1)\varepsilon_{m}E_{l}+C_{l}(l+1)\Big[\varepsilon_{s}(\omega)-\varepsilon_{m}\Big]}
{l\varepsilon_{s}(\omega)+(l+1)\varepsilon_{m}}.
\end{equation}
Now we substitute $B_{l}$ for $B_{n}$ (by changing $l$ to $n$) in Eq. (\ref{e38}), and rearrange terms to finally obtain
\begin{equation}\label{e40}
-\sum_{n=1}^{N}K_{ln}E_{n}z_{n} = \sum_{n=1}^{N}K_{ln}C_{n}y_{n}+x_{l}\sum_{n=1}^{N}M_{ln}C_{n},~~ l = 1,2,...,N,
\end{equation}
where 
\begin{eqnarray}
x_{l} \equiv 
\frac{\Big[l\varepsilon_{c}+(l+1)\varepsilon_{s}(\omega)\Big]}{l\Big[\varepsilon_{c}-\varepsilon_{s}(\omega)\Big]}, \label{e41}\\
y_{n} \equiv \frac{(n+1)\Big[\varepsilon_{s}(\omega)-\varepsilon_{m}\Big]}{\Big[n\varepsilon_{s}(\omega)+(n+1)\varepsilon_{m}\Big]}, \label{e42}\\
z_{n} \equiv \frac{(2n+1)\varepsilon_{m}}{\Big[n\varepsilon_{s}(\omega)+(n+1)\varepsilon_{m}\Big]}. \label{e43}
\end{eqnarray}
Eq. (\ref{e40}) forms a system of $N\times N$ simultaneous linear equations with $N$ unknowns, where the $C_{n}$ terms are the unknowns, 
since $E_{n}$ is given by Eq. (\ref{e6}). Given the necessary input parameters, we have written a python code that solves Eq. (\ref{e40})
for $N = 15$, which was numerically sufficient to accurately demonstrate the multipolar contributions based on the MNP size we considered. 

Note that Eq. (\ref{e37}) can be re-written as 
\begin{equation}\label{e44}
D_{l} = -\tilde{\alpha}_{l}(\omega) E_{l}, 
\end{equation}
where $\tilde{\alpha}_{l}(\omega)$ is a $b^{2l+1}$-normalized static multipole polarizability of the nanoegg, which arises 
because we normalized $r_{c}$ and $r_{s}$ with their respective values at the boundaries.  
By comparing Eq. (\ref{e44}) to Eq. (\ref{e37}), we obtain 
\begin{equation}\label{e45}
\tilde{\alpha}_{l}(\omega) = 
\frac{ (2l+1)\varepsilon_{s}(\omega)\Big[\frac{C_{l}}{-E_{l}}\Big]+l\Big[\varepsilon_{s}(\omega)-\varepsilon_{m}\Big] }
{ l\varepsilon_{s}(\omega)+(l+1)\varepsilon_{m} },
\end{equation}
so that by substituting Eq. (\ref{e6}) and Eq. (\ref{e44}) into Eq. (\ref{e7}), we obtain
\begin{equation}\label{e46}
\Phi_{ind}(r_{s},\theta_{s}) = \sum_{n=1}^{\infty}\alpha_{n}(\omega)\frac{\text{p}_{z}(n+1)}{\varepsilon_{m} r^{n+2}}
\Big(\frac{1}{r_{s}}\Big)^{n+1}P_{n}(\cos\theta_{s}),
\end{equation}
where 
\begin{equation}\label{e47}
\alpha_{n}(\omega) = b^{2n+1}\tilde{\alpha}_{n}(\omega), 
\end{equation}
is the static multipole polarizability of the nanoegg. 
From Eq. (\ref{e46}), we obtain the induced dipole potential as
\begin{equation}\label{e48}
\Phi_{ind}(r_{s},\theta_{s})\Big|_{n=1} = \frac{\text{p}_{z,ind}\cos\theta_{s}}{\varepsilon_{m} r_{s}^{2}},
\end{equation}
so that the induced dipole moment in the case of perpendicular dipole orientation is
\begin{equation}\label{e49}
\text{p}_{z,ind} = \alpha_{1}(\omega)\frac{2}{r^3}\text{p}_{z}.
\end{equation}
Next, we calculate the induced multipolar field through 
$\mathbf{E}_{z,ind}(r_{s},\theta_{s}) = -\nabla\Phi_{z,ind}(r_{s},\theta_{s})$ as follows
\begin{eqnarray}
\mathbf{E}_{z,ind}(r_{s},\theta_{s}) = -\dfrac{\partial\Phi_{ind}}{\partial r_{s}}\hat{r}_{s}-
\frac{1}{r_{s}}\dfrac{\partial\Phi_{ind}}{\partial \theta_{s}}\hat{\theta}_{s},
~~~~~~~~~~~~~~~~~~~~~~~~~~~~~~~~~~~~~~~~~~~~~~~~~~~~~~~~~~~~~~~~\nonumber\\
= \sum_{n=1}^{\infty}\alpha_{n}(\omega)\frac{\text{p}_{z}(n+1)}{\varepsilon_{m} r_{s}^{n+2}r^{n+2}}\left[
(n+1)P_{n}(\cos\theta_{s})\hat{r}_{s}-
\dfrac{d}{d\theta_{s}}\Big[P_{n}(\cos\theta_{s})\Big]\hat{\theta}_{s} \right],
\end{eqnarray}
and using the properties of the Legendre function of the first kind: 
\begin{equation}\nonumber
\frac{P_{n}(\cos\theta_{s})}{\cos\theta_{s}}\Big|_{\theta_{s}\rightarrow 0} = 1, ~~
\dfrac{d}{d\theta_{s}}\Big[P_{n}(\cos\theta_{s})\Big]\Big|_{\theta_{s}\rightarrow 0} = 0,
\end{equation}
we obtain the induced field at the dipole position $r_{s} = r$ and $\theta_{s}\rightarrow 0$ as
\begin{equation}\label{e51}
\mathbf{E}_{z,ind}(r_{s} = r,\theta_{s}\rightarrow 0) = 
\sum_{n=1}^{\infty}\alpha_{n}(\omega)\frac{\text{p}_{z}(n+1)^{2}}{\varepsilon_{m} r^{2n+4}}\hat{z},
\end{equation}
with 
\begin{equation}\label{e52}
\hat{z} = 
\cos\theta_{s}\hat{r}_{s}.
%-0\sin\theta_{s}\hat{\theta}_{s}.
\end{equation}

We can now derive expressions for parameters that define the optical response of the molecular dipole at the emission stage. We begin with 
the Purcell factor. The radiative decay rate of the molecular dipole in the presence of the MNP and the embedding medium, normalized by the radiative decay rate of the molecule in the same medium in the absence of the MNP is known as Purcell factor \cite{BN07,Khat14,Chris16}. 
It is defined as \cite{BN07}
\begin{equation}\label{e53}
P = \frac{|\mathbf{p}+\mathbf{p}_{ind}|^{2}}{|\mathbf{p}|^{2}}.
\end{equation}
Hence, for the perpendicular dipole orientation, we substitute Eq. (\ref{e49}) into Eq. (\ref{e53}) to obtain the Purcell factor as
\begin{equation}\label{e54}
P_{\perp} = \left|1 + \left[\frac{\alpha_{1}(\omega)}{1-2i\alpha_{1}(\omega)\frac{k^{3}}{3}}\right]\frac{2}{r^3}\right|^{2},
\end{equation}
where the static dipole polarizability has been corrected for radiation damping following the prescription in Refs. \cite{FW84,Mert07}. 
$k = 2\pi\sqrt{\varepsilon_{m}}/\lambda$ is the wavenumber of light in the medium, and $\lambda$ is the emission wavelength. 

The rate of power dissipation by the molecular dipole near the MNP surrounded by a dielectric medium, 
normalized by the radiative decay rate of the dipole in the same medium in the absence of the MNP is given by \cite{FW84,Moroz11} 
\begin{equation}\label{e55}
\frac{\gamma_{diss}}{\gamma^{o}_{rad}} = \frac{3\varepsilon_{m}}{2k^{3}}\frac{\Im[\mathbf{p^{*}.E}_{ind}]}{|\mathbf{p}|^{2}}. 
\end{equation}
Substituting Eq. (\ref{e51}) into Eq. (\ref{e55}), we obtain the normalized energy transfer rate in the case of a perpendicular dipole 
as 
\begin{equation}\label{e56}
\frac{\gamma_{diss,\perp}}{\gamma^{o}_{rad}}  = \frac{3}{2(kr)^{3}}\left[\frac{4}{r^{3}}\Im\left[\frac{\alpha_{1}(\omega)}{1-2i\alpha_{1}(\omega)\frac{k^{3}}{3}}\right] + 
\sum_{n=2}^{N}\frac{(n+1)^{2}}{r^{2n+1}}\Im[\alpha_{n}(\omega)]\right]. 
\end{equation}

Energy conservation requires that the rate of power dissipation by the excited molecule must be equal to the non-radiative energy transfer rate
from the molecule to the MNP. This energy is then absorbed by the MNP where it is dissipated as heat due to Ohmic heating in the metal \cite{GN81,FW84,Rupp82,Chris16}. Also, if we consider the two-way antenna model described in Ref. \cite{BN07}, where 
the molecular dipole acts as a transmitter of radiation while the MNP acts as a receiver or vice-versa, then the Purcell factor can also be 
regarded as the normalized rate of radiative energy transfer from the MNP to the molecular dipole. In all, we have 
\begin{eqnarray}
\text{(Emitter)}~~~\frac{\gamma_{diss}}{\gamma^{o}_{rad}} = \frac{\gamma_{abs}}{\gamma^{o}_{rad}}~~~\text{(MNP)}, \label{e57}\\
P = \frac{\gamma_{rad}}{\gamma^{o}_{rad}}.~~~~~~~~~~~~~ \label{e58}
\end{eqnarray}

We can now calculate the antenna efficiency, also known as the quantum yield of the antenna, when the dipole is normal to the MNP surface, using \cite{GN81}
\begin{equation}\label{e59}
\eta_{\perp} = \frac{\gamma_{rad,\perp}}{\gamma^{o}_{rad}}\left[\frac{\gamma_{rad,\perp}}{\gamma^{o}_{rad}}
+\frac{\gamma_{abs,\perp}}{\gamma^{o}_{rad}}\right]^{-1}.
\end{equation}

Finally, we derive an expression for the quantum yield of the molecule in the presence of the MNP. Let $Y_{o}$ denote the 
intrinsic quantum yield of the molecule, defined as \cite{BN07}
\begin{equation}\label{e60}
Y_{o} = \frac{\gamma^{o}_{rad}}{\gamma^{o}_{rad}+\gamma^{o}_{nrad}}, 
\end{equation}
where $\gamma^{o}_{rad}$ and $\gamma^{o}_{nrad}$ are the intrinsic radiative and non-radiative decay rates of the excited molecule. 
Since the coupling of the dipole field of the excited molecule to plasmonic modes of the MNP does not affect the intrinsic non-radiative 
decay rate \cite{BN07,Wien14,Chris16}, the modified quantum yield is defined in terms of the Purcell factor and the total decay rate 
as \cite{BN07,Khat14}
\begin{equation}\label{e61}
Y = P\left[P+\frac{\gamma_{diss}}{\gamma^{o}_{rad}}+\frac{\gamma^{o}_{nrad}}{\gamma^{o}_{rad}}\right]^{-1}.  
\end{equation}
After eliminating $\gamma^{o}_{nrad}/\gamma^{o}_{rad}$ from Eq. (\ref{e61}) using Eq. (\ref{e60}), as well as $\gamma_{diss}/\gamma^{o}_{rad}$ using Eq. (\ref{e59}), making use of Eqs. (\ref{e57}) and (\ref{e58}), with some re-arrangement, we obtain the well-known formula for
quantum yield enhancement as \cite{Moha08,Wien14}
\begin{equation}\label{e62}
Y = Y_{o}\left[\frac{Y_{o}}{\eta} + \frac{1-Y_{o}}{P}\right]^{-1},
\end{equation}
so that for the perpendicular dipole, $\eta = \eta_{\perp}$ and $P = P_{\perp}$.

\subsection{Parallel Dipole}\label{s2.2}
When the molecular dipole is tangential to the surface of the MNP, both the dipole potential and the electric potentials in the core, shell,
and medium regions of the MNP, are dependent on the azimuth angle $\phi$ of the dipole \cite{FW84,Maj17}. For the 
molecule-nanoegg system, these potentials can therefore be written as \cite{FW84,Maj17}
\begin{eqnarray}
\Phi_{c}(r_{c},\theta_{c}) = \sum_{n=1}^{\infty}A_{n}\Big(\frac{r_{c}}{a}\Big)^{n}P^{1}_{n}(\cos\theta_{c})\cos\phi,\label{e63}\\
\Phi_{s}(r_{s},\theta_{s}) = \sum_{n=1}^{\infty}\Big[B_{n}\Big(\frac{r_{s}}{b}\Big)^{n}+
C_{n}\Big(\frac{b}{r_{s}}\Big)^{n+1}\Big]P^{1}_{n}(\cos\theta_{s})\cos\phi, \label{e64}\\
\Phi_{m}(r_{s},\theta_{s}) = \Phi_{dip}(r_{p},\theta_{p})+\Phi_{ind}(r_{s},\theta_{s}), \label{e65}
\end{eqnarray}
where \cite{FW84,Maj17}
\begin{equation}\label{e66}
\Phi_{dip}(r_{p},\theta_{p}) = -\frac{\mathbf{p}.\mathbf{x}}{\varepsilon_{m}r^{3}_{p}} = 
-\frac{\text{p}_{x}\sin\theta_{p}\cos\phi}{\varepsilon_{m}r^{2}_{p}} = 
\sum_{n=1}^{\infty}E_{n}\Big(\frac{r_{s}}{b}\Big)^{n}P^{1}_{n}(\cos\theta_{s})\cos\phi,
\end{equation}
with 
\begin{equation}\label{e67}
E_{n} = \frac{\text{p}_{x}b^{n}}{\varepsilon_{m} r^{n+2}}, r = b+d,
\end{equation}
and 
\begin{equation}\label{e68}
\Phi_{ind}(r_{s},\theta_{s}) = \sum_{n=1}^{\infty}D_{n}\Big(\frac{b}{r_{s}}\Big)^{n+1}P^{1}_{n}(\cos\theta_{s})\cos\phi.
\end{equation}
Here, $r_{c}$ and $r_{s}$ have been normalized with their respective values at the boundaries, and $P^{1}_{n}(u)$ is the 
associated Legendre function of the first kind, evaluated at $m = 1$, which corresponds to the parallel orientation of the dipole.
$m$ is the azimuthal number.
$A_{n}, B_{n}$ and $C_{n}$, and $D_{n}$ are the complex amplitudes of the electrostatic potential in the core, shell, medium regions 
of the nanoegg-emitter system respectively, for the tangential dipole. $E_{n}$ is the amplitude of the source tangential dipole potential.

At the boundaries, both the potential and the normal component of the displacement field must be continuous, leading to
the following boundary conditions \cite{Nort16}: 
\begin{eqnarray}
\Phi_{c}\Big(r_{c},\theta_{c}\Big)\Big|_{r_{c}=a}
= \Phi_{s}\Big(r_{s},\theta_{s}\Big)\Big|_{r_{c}=a}, \label{e69}\\
\Phi_{s}\Big(r_{s},\theta_{s}\Big)\Big|_{r_{s} = b} = \Phi_{m}\Big(r_{s},\theta_{s}\Big)\Big|_{r_{s} = b}, \label{e70}\\
\varepsilon_{c}\dfrac{\partial \Phi_{c}\Big(r_{c},\theta_{c}\Big)}{\partial r_{c}}\Big|_{r_{c} = a} = 
\varepsilon_{s}(\omega)\dfrac{\partial \Phi_{s}\Big(r_{s},\theta_{s}\Big)}{\partial r_{c}}\Big|_{r_{c} = a},\label{e71}\\
\varepsilon_{s}(\omega)\dfrac{\partial \Phi_{s}\Big(r_{s},\theta_{s}\Big)}{\partial r_{s}}\Big|_{r_{s} = b} = 
\varepsilon_{m}\dfrac{\partial \Phi_{m}\Big(r_{s},\theta_{s}\Big)}{\partial r_{s}}\Big|_{r_{s} = b}. \label{e72}
\end{eqnarray}
Setting $u_{c} \equiv \cos\theta_{c}$ and $u_{s} \equiv \cos\theta_{s}$, and 
combining Eqs. (\ref{e63}-\ref{e68}) and Eqs. (\ref{e69}-\ref{e72}), we obtain:
\begin{eqnarray}
\sum_{n=1}^{\infty}A_{n}P^{1}_{n}(u_{c}) = \sum_{n=1}^{\infty}\Big[B_{n}\Big(\frac{r_{s}}{b}\Big)^{n}+
C_{n}\Big(\frac{b}{r_{s}}\Big)^{(n+1)}\Big]P^{1}_{n}(u_{s})\Big|_{r_{c} = a}, \label{e73}\\
\sum_{n=1}^{\infty}[B_{n}+C_{n}]P^{1}_{n}(u_{s}) = \sum_{n=1}^{\infty}[E_{n}+D_{n}]P^{1}_{n}(u_{s}), \label{e74}\\
\varepsilon_{c}\sum_{n=1}^{\infty}nA_{n}P^{1}_{n}(u_{c}) = 
a\varepsilon_{s}(\omega)\sum_{n=1}^{\infty} \dfrac{\partial}{\partial r_{c}} \Big[B_{n}\Big(\frac{r_{s}}{b}\Big)^{n}+
C_{n}\Big(\frac{b}{r_{s}}\Big)^{(n+1)}\Big]P^{1}_{n}(u_{s})\Big|_{r_{c} = a}, \label{e75}\\
\varepsilon_{s}(\omega)\sum_{n=1}^{\infty}[nB_{n}-(n+1)C_{n}]P^{1}_{n}(u_{s}) = 
\varepsilon_{m}\sum_{n=1}^{\infty}[nE_{n}-(n+1)D_{n}]P^{1}_{n}(u_{s}). \label{e76}
\end{eqnarray}
Multiplying both sides of Eqs. (\ref{e73} \& \ref{e75}) each by $P^{1}_{l}(u_{c})$
and  Eqs. (\ref{e74} \& \ref{e76}) each by $P^{1}_{l}(u_{s})$, and integrating each one respectively via 
\begin{equation}\label{e77}
\rho_{l} \int_{-1}^{1}P^{1}_{l}(u)P^{1}_{n}(u)du = \delta_{ln}, ~~\rho_{l} = \frac{2l+1}{2l(l+1)},
\end{equation}
we obtain 
\begin{eqnarray}
A_{l} = \sum_{n=1}^{\infty}K_{ln}B_{n}+\sum_{n=1}^{\infty}M_{ln}C_{n}, \label{e78}\\
B_{l}+C_{l} = E_{l}+D_{l}, \label{e79}\\
\varepsilon_{c}lA_{l} = \varepsilon_{s}(\omega)\Big[\sum_{n=1}^{\infty}L_{ln}B_{n}+\sum_{n=1}^{\infty}N_{ln}C_{n}
\Big], \label{e80}\\
\varepsilon_{s}(\omega)[lB_{l}-(l+1)C_{l}] = 
\varepsilon_{m}[lE_{l}-(l+1)D_{l}], \label{e81}
\end{eqnarray}
where 
\begin{eqnarray}
K_{ln} = \frac{\rho_{l}}{b^{n}}\int_{-1}^{1}r_{s}^{n}P^{1}_{n}(u_{s})\Big|_{r_{c} = a}P^{1}_{l}(u_{c})du_{c}, \label{e82}\\
M_{ln} = \rho_{l}b^{n+1}\int_{-1}^{1}\frac{P^{1}_{n}(u_{s})}{r_{s}^{n+1}}\Big|_{r_{c}=a}P^{1}_{l}(u_{c})du_{c}, \label{e83}\\
L_{ln} = \frac{\rho_{l}a}{b^{n}}\int_{-1}^{1}
\dfrac{\partial }{\partial r_{c}}\Big[r_{s}^{n}P^{1}_{n}(u_{s})\Big]_{r_{c} = a}P^{1}_{l}(u_{c})du_{c}, \label{e84}\\
N_{ln} = \rho_{l}ab^{n+1}\int_{-1}^{1}
\dfrac{\partial }{\partial r_{c}}\left[\frac{P^{1}_{n}(u_{s})}{r_{s}^{n+1}}\right]_{r_{c}=a}P^{1}_{l}(u_{c})du_{c}. \label{e85}
\end{eqnarray}

Since the static polarizability of a spherical MNP remains the same in both the perpendicular and parallel dipole orientations 
\cite{FW84,Maj17}, we assume that the interior and exterior solutions of the Laplace equation for the parallel dipole
also obey the SHAT in a similar manner as that of the perpendicular dipole (although for $m=1$, the SHAT is slightly different, see Ref. \cite{Cala78}), so that the polarizability of the nanoegg remains the same. 
Hence, the values of Eqs. (\ref{e82}-\ref{e85}) are the same as those of the perpendicular case, 
and Eqs. (\ref{e37}-\ref{e45}) are also retained. However, the coefficients $E_{l}$ are now given by Eq. (\ref{e67}), and the induced potential is different in this case. 

By substituting Eq. (\ref{e67}) and Eq. (\ref{e44}) into Eq. (\ref{e68}), we obtain
\begin{equation}\label{e86}
\Phi_{ind}(r_{s},\theta_{s}) = -\sum_{n=1}^{\infty}\alpha_{n}(\omega)\frac{\text{p}_{x}}{\varepsilon_{m} r^{n+2}}
\Big(\frac{1}{r_{s}}\Big)^{n+1}P^{1}_{n}(\cos\theta_{s})\cos\phi,
\end{equation}
From Eq. (\ref{e86}), we obtain the induced dipole potential as
\begin{equation}\label{e87}
\Phi_{ind}(r_{s},\theta_{s})\Big|_{n=1} = -\frac{\text{p}_{x,ind}\sin\theta_{s}\cos\phi}{\varepsilon_{m} r_{s}^{2}},
\end{equation}
so that the induced dipole moment in the case of parallel dipole orientation is
\begin{equation}\label{e88}
\text{p}_{x,ind} = -\alpha_{1}(\omega)\frac{1}{r^3}\text{p}_{x}.
\end{equation}
We substitute Eq. (\ref{e88}) into Eq. (\ref{e53}) to obtain the Purcell factor for the parallel dipole as
\begin{equation}\label{e89}
P_{\parallel} = \left|1 - \left[\frac{\alpha_{1}(\omega)}{1-2i\alpha_{1}(\omega)\frac{k^{3}}{3}}\right]\frac{1}{r^3}\right|^{2}.
\end{equation}
Again, we have corrected the dipole polarizability for radiation reaction.

The induced multipolar field in this case is calculated via 
$\mathbf{E}_{x,ind}(r_{s},\theta_{s}) = -\nabla\Phi_{x,ind}(r_{s},\theta_{s})$ as follows
\begin{equation}
\begin{aligned}
\mathbf{E}_{x,ind}(r_{s},\theta_{s}) = {} & -\dfrac{\partial\Phi_{ind}}{\partial r_{s}}\hat{r}_{s}-
\frac{1}{r_{s}}\dfrac{\partial\Phi_{ind}}{\partial \theta_{s}}\hat{\theta}_{s}-
\frac{1}{r_{s}\sin\theta_{s}}\dfrac{\partial\Phi_{ind}}{\partial \phi}\hat{\phi}_{s}\\
= {} & \sum_{n=1}^{\infty}\alpha_{n}(\omega)\frac{\text{p}_{x}}{\varepsilon_{m} r_{s}^{n+2}r^{n+2}}\Big[-
(n+1)P^{1}_{n}(\cos\theta_{s})\cos\phi\hat{r}_{s}\\
{} & + \dfrac{d}{d\theta_{s}}\Big[P^{1}_{n}(\cos\theta_{s})\Big]\cos\phi\hat{\theta}_{s} -
\frac{P^{1}_{n}(\cos\theta_{s})}{\sin\theta_{s}}\sin\phi\hat{\phi}_{s}\Big],~~~~~~~~\label{e90}
\end{aligned}
\end{equation}
and using the properties of the associated Legendre function of the first kind:
\begin{equation}\label{e91}
P^{1}_{n}(\cos\theta_{s})\Big|_{\theta_{s}\rightarrow 0} = 0,~~
\frac{1}{\cos\theta_{s}}\dfrac{d}{d\theta_{s}}\Big[P^{1}_{n}(\cos\theta_{s})\Big]\Big|_{\theta_{s}\rightarrow 0} =\frac{n(n+1)}{2},~~
\frac{P^{1}_{n}(\cos\theta_{s})}{\sin\theta_{s}}\Big|_{\theta_{s}\rightarrow 0} =\frac{n(n+1)}{2},
\end{equation}
we obtain the induced field at the dipole position $r_{s} = r$ and $\theta_{s}\rightarrow 0$ as
\begin{equation}\label{e92}
\mathbf{E}_{x,ind}(r_{s} = r,\theta_{s}\rightarrow 0) = 
\sum_{n=1}^{\infty}\alpha_{n}(\omega)\frac{\text{p}_{x}n(n+1)}{2\varepsilon_{m} r^{2n+4}}\hat{x},
\end{equation}
with 
\begin{equation}\label{e93}
\hat{x} = %0sin\theta_{s}cos\phi\hat{r}_{s}+
\cos\theta_{s}\cos\phi\hat{\theta}_{s}-\sin\phi\hat{\phi}.
\end{equation}
Substituting Eq. (\ref{e92}) into Eq. (\ref{e57}), we obtain the normalized rate of power dissipation by the parallel dipole 
as 
\begin{equation}\label{e94}
\frac{\gamma_{diss,\parallel}}{\gamma^{o}_{rad}} = \frac{3}{4(kr)^{3}}\left[\frac{2}{r^{3}}\Im\left[\frac{\alpha_{1}(\omega)}{1-2i\alpha_{1}(\omega)\frac{k^{3}}{3}}\right] + 
\sum_{n=2}^{N}\frac{n(n+1)}{r^{2n+1}}\Im[\alpha_{n}(\omega)]\right]. 
\end{equation}
Eqs. (\ref{e92} \& \ref{e57}) also hold for the parallel dipole, as well as Eqs. (\ref{e59} \& \ref{e62}) 
with $P = P_{\parallel}$ and $\eta = \eta_{\parallel}$.

\section{Results and Discussion}\label{s3}
We consider a DCMS nanoegg with dimensions $a = 15$ nm, $b = 20$ nm, a silica core of dielectric constant $\varepsilon_{c} = 2.13$ \cite{Chris16}, for the following core-offsets $\sigma = 0.0, 0.5, 1.0, 1.5, $ and $2.0$ nm, and a gold shell with the local dielectric function given in Eq. (\ref{e1}). The MNP size we have chosen allows us to discuss the enhancement factors of a weak emitter such as crystal violet (CV) near the nanoegg. CV molecules have an intrinsic quantum yield of $Y_{o} = 2\%$, an intrinsic radiative decay rate $\gamma^{o}_{rad} = 1.9\times10^{7}$ s$^{-1}$, and a peak emission wavelength of $640$ nm when excited at $633$ nm \cite{Khat14}. 

PEF or emission rate enhancement is calculated via \cite{BN07,Moha08,Beyer11,Khat14,Chris16,Wei17}
\begin{equation}\label{e95}
\frac{\gamma_{em}}{\gamma^{o}_{em}}(\omega_{exc},\omega_{em})  = 
\frac{\gamma_{exc}}{\gamma^{o}_{exc}}(\omega_{exc}) \frac{Y}{Y_{o}}(\omega_{em}), 
\end{equation}
where $\gamma_{exc}/\gamma^{o}_{exc}$ is excitation rate enhancement of the molecule evaluated at the excitation frequency, 
$\omega_{exc}$, and $Y$ is the quantum yield enhancement of the molecule evaluated at the peak emission frequency, 
$\omega_{em}$. 

As mentioned earlier, we will consider the emission stage of PEF in detail, and mention the overall stage i.e excitation and emission,
towards the end of this section. 
The optimal range of MNP-molecule distance for PEF has been reported as $\sim 3-5$ nm for nanorods \cite{Khat14,Wien14},
$\sim 2-3$ nm for nanospheres \cite{Anger06,Chow09,Mert07}, and $\sim 2-7$ nm for nanoshells \cite{Bard08,Bard09,Chris16}. We have chosen to use a fixed MNP-molecule distance of $d = 5$ nm. The MNP-molecule system is surrounded by water which has a dielectric constant $\varepsilon_{m} \approx 1.78$.
\begin{figure}[h!] 
	\centering
	\includegraphics[width = 0.697\linewidth]{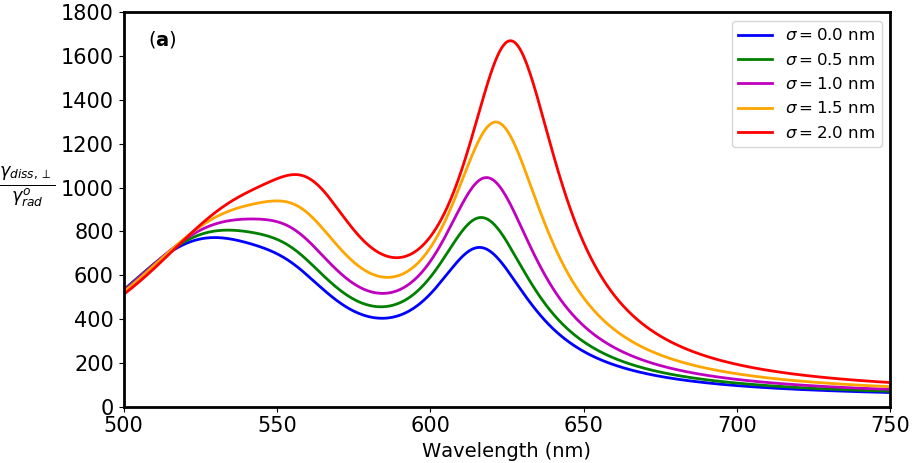}\\
	\includegraphics[width = 0.7\linewidth]{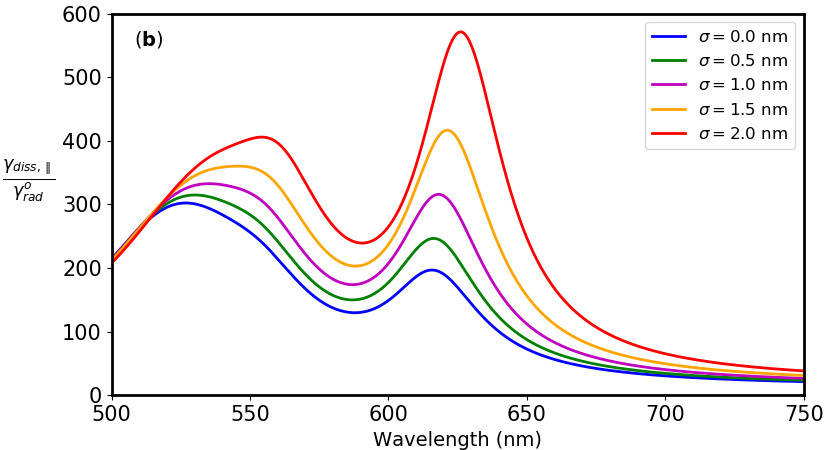}
	\caption{\small{Dependence of the normalized rate of power dissipation by the molecular dipole on the core-offset of the
			nanoegg in the case of (\textbf{a}) perpendicular dipole and (\textbf{b}) parallel dipole orientations of the molecule near the nanoegg. 
	}} 
	\label{f2}
\end{figure}

Figs. \ref{f2}(\textbf{a}) and (\textbf{b}) present the normalized rate of non-radiative energy transfer from an excited molecule at $d = 5$ nm for the normal and tangential dipole orientations, respectively. In both plots, the dipolar LSPR undergoes a redshift from $616$ nm at $\sigma=0.0$ nm to $626$ nm at $\sigma=2.0$ nm. 
For the normal dipole, Fig. \ref{f2}(\textbf{a}), at $\sigma>0.0$ nm, the non-radiative energy transfer rate reaches a maximum at the dipolar LSPR of the nanoegg, while this occurs for the tangential dipole at $\sigma>1.0$ nm, Fig. \ref{f2}(\textbf{b}). 
This is because the dipolar near-field of the excited molecule is most strongly coupled to the 
dipolar surface plasmon mode of the nanoegg. A similar behaviour has been reported for spheres \cite{Rupp82} and nanorods \cite{Khat14}. 
However, the energy transfer rate of the normal dipole is more than twice that of the tangential dipole, for the same emission wavelength. 
The dissipative, blue-shifted peaks in Figs. \ref{f2}(\textbf{a}) and (\textbf{b}) are due to the coupling of the dipole field 
of the excited molecule to higher-order ($l\ge2$) surface plasmon modes of the nanoegg. The impact of the dipole-active modes on the 
energy transfer rate can be seen in the dramatic increase in the peaks as the core is off-set from $\sigma=0.0$ nm to $\sigma=2.0$ nm.
For a CV molecule at $d = 5$ nm from the nanoegg, the non-radiative energy transfer rate will therefore increase from $\sim 400\times\gamma^{o}_{rad}$ at $\sigma=0.0$ nm to nearly $1500\times\gamma^{o}_{rad}$ at $\sigma=2.0$ nm, for the normal dipole (Fig. \ref{f2}(\textbf{a}) at $\lambda = 640$ nm), and from $\sim 100\times\gamma^{o}_{rad}$ at $\sigma=0.0$ nm to nearly $400\times\gamma^{o}_{rad}$ at $\sigma=2.0$ nm, for the tangential dipole (Fig. \ref{f2}(\textbf{b}) at $\lambda = 640$ nm).
\begin{figure}[h!]
	\centering
	\includegraphics[width = 0.54\linewidth]{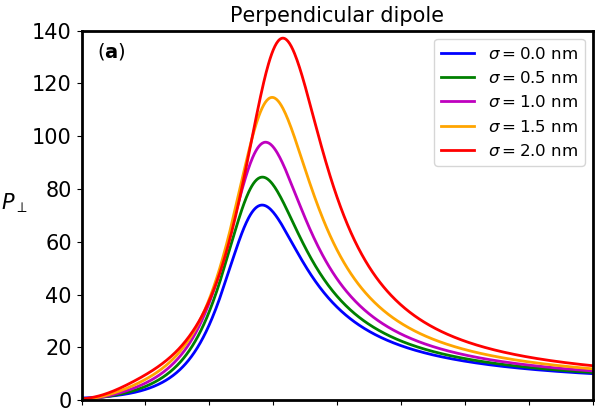}\\
	\includegraphics[width = 0.54\linewidth]{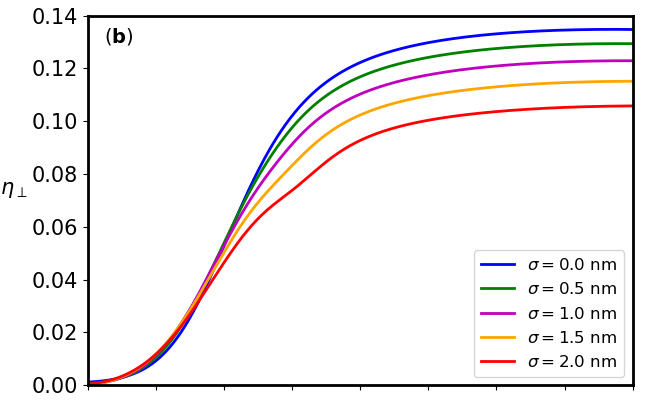}\\
	\includegraphics[width = 0.54\linewidth]{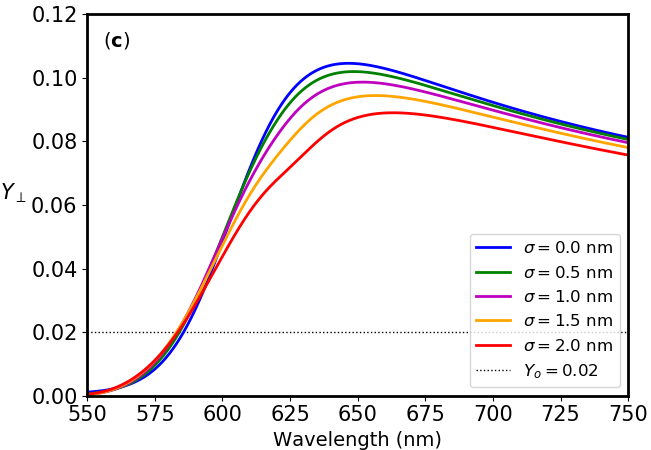}
	\caption{\small{Dependence of the (\textbf{a}) Purcell factor, (\textbf{b}) antenna efficiency, and (\textbf{c}) quantum yield of the molecular dipole on the core-offset of the nanoegg in the case of perpendicular dipole orientation. 
	}} 
	\label{f3}
\end{figure}

\begin{figure}[h!]
	\centering
	\includegraphics[width = 0.52\linewidth]{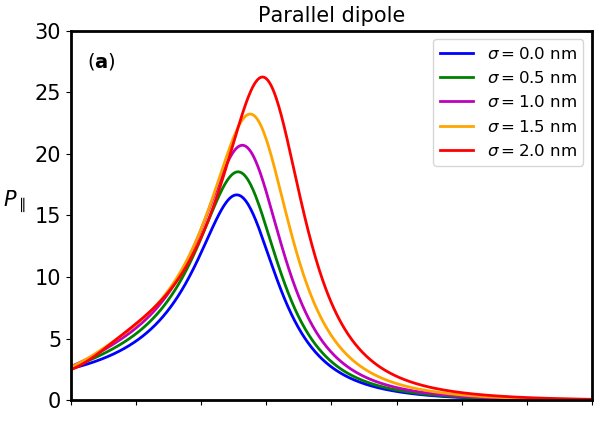}\\
	\includegraphics[width = 0.54\linewidth]{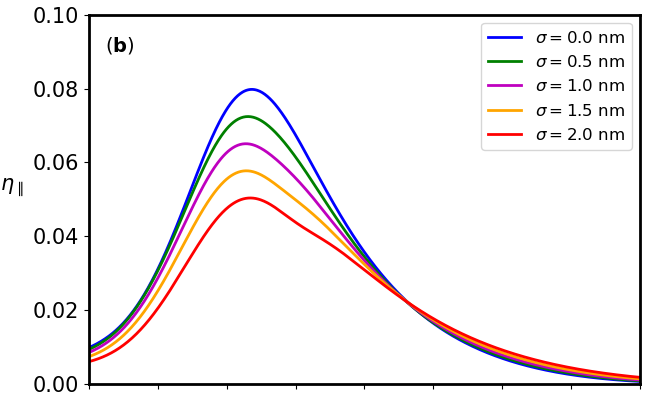}\\
	~\includegraphics[width = 0.54\linewidth]{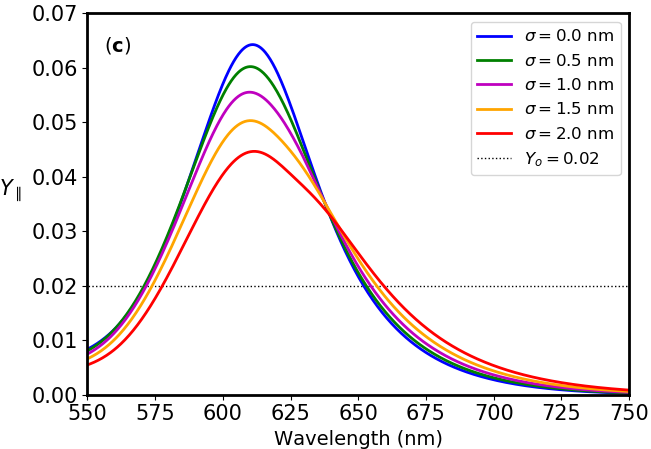}
	\caption{\small{Dependence of the (\textbf{a}) Purcell factor, (\textbf{b}) antenna efficiency, and (\textbf{c}) quantum yield of the molecular dipole on the core-offset of the nanoegg in the case of parallel dipole orientation. 
	}} 
	\label{f4} 
\end{figure}

In Figs. \ref{f3}(\textbf{a}) and \ref{f4}(\textbf{a}), the Purcell factors of an excited molecular dipole at $d = 5$ nm from the nanoegg 
are shown for the normal and tangential dipole respectively. As the core-offset increases from $\sigma=0.0$ nm to $\sigma=2.0$ nm, both plots show a redshift in the peak emission wavelength at which radiative decay rate enhancement occurs. 
In comparison to Figs. \ref{f2}(\textbf{a}) and (\textbf{b}), the peak emission wavelengths for the 
normal dipole are redshifted from the dipolar LSPR, while the peak emission wavelengths for the tangential dipole are blue-shifted from 
the dipolar LSPR. The intrinsic radiative decay rate of the molecule is more enhanced for the normal dipole because the induced dipole moment is stronger in the normal orientation of the dipole. 
For a CV molecule at $d = 5$ nm from the nanoegg, the radiative decay rate will therefore increase from $\sim 40\times\gamma^{o}_{rad}$ at $\sigma=0.0$ nm to nearly $130\times\gamma^{o}_{rad}$ at $\sigma=2.0$ nm, for the normal dipole (Fig. \ref{f3}(\textbf{a}) at $\lambda = 640$ nm), and from $\sim 4\times\gamma^{o}_{rad}$ at $\sigma=0.0$ nm to nearly $25\times\gamma^{o}_{rad}$ at $\sigma=2.0$ nm, for the tangential dipole (Fig. \ref{f4}(\textbf{a}) at $\lambda = 640$ nm).
 
When embedded in water, the silica core-gold nanoshell is a passive antenna below $550$ nm and an active antenna above $550$ nm. 
This is due to the dependence of the radiative power on the bright mode of the nanoshell, only.
The antenna is more efficient in the normal dipole orientation (Fig. \ref{f3}(\textbf{b})) than in the tangential dipole 
orientation (Fig. \ref{f4}(\textbf{b})). This is because a stronger incident dipole field reaches the nanoegg in the normal orientation, 
causing the antenna to radiate more power. Due to a much higher increase in the absorbed power compared to the increase in radiative power of the nanoegg, as the core-offset increases, the antenna efficiency decreases with increasing core-offset, regardless of the dipole orientation.
Beyond the peak emission wavelength, the antenna efficiency plateaus for the normal dipole, because both the radiative and absorptive powers 
tend towards constant values, while for the tangential dipole, the absorptive power continues to dominate the radiative power.

For the normal dipole, the peak values of the modified quantum yield of the molecule occur at emission wavelengths red-shifted from
the dipolar LSPR of the nanoegg (Fig. \ref{f3} (\textbf{c})), while those of the tangential dipole occur at emission wavelengths blue-shifted from the dipolar LSPR (Fig. \ref{f4} (\textbf{c})), due to the different contributions from the Purcell factor. 
For a given emission wavelength and MNP-molecule distance, the modified quantum yield of the molecule is always less than the antenna 
efficiency due to the contribution of the intrinsic non-radiative decay rate of the molecule to the total decay rate of the MNP-molecule
system.

Although the intrinsic radiative decay rate of the excited molecule is more enhanced near the nanoegg, its intrinsic quantum yield is 
less enhanced when compared to those of the same molecule near a concentric nanoshell. This is due to the high rate of non-radiative 
energy transfer from the excited molecule to the nanoegg compared to the nanoshell.  
The intrinsic quantum yield of the CV molecule emitting at $640$ nm increases from $2$\% to $\sim 10$\% at $\sigma = 0$ nm and from 
$2$\% to $\sim 8$\% at $\sigma = 2.0$ nm for the normal dipole, and from $2$\% to $\sim 3.3$\% at $\sigma = 0$ nm and from 
$2$\% to $3$\% at $\sigma = 2.0$ nm for the tangential dipole.

\begin{figure}[h!] 
	\centering
	\includegraphics[width = 0.60\linewidth]{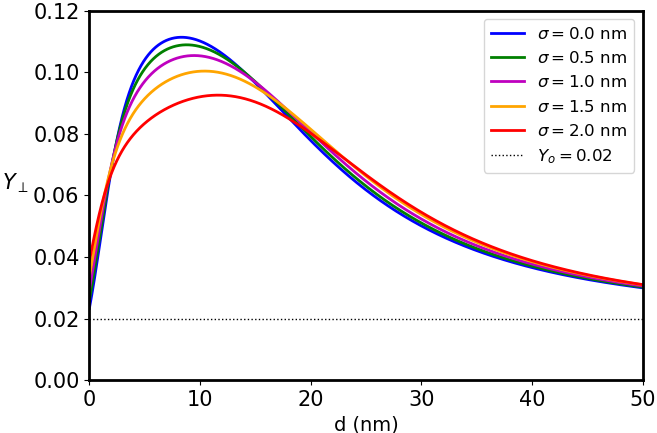}
	\caption{\small{Quantum yield enhancement of a CV molecule at the peak emission wavelength $\lambda = 640$ nm, for the core-offsets
			studied, as a function of its distance from the nanoegg, for a perpendicularly-oriented molecular dipole. 
	}} 
	\label{f5}
\end{figure}

Fig. \ref{f5} shows that for the CV molecule, the quenching of the intrinsic quantum yield which occurs when $Y<Y_{o}$, is very unlikely, 
even at short distances ($d\longrightarrow 0$) where the dissipation rate is very high.  This is due to the 
low value of $Y_{o}$, so that $\eta$ is always greater than or equal to $Y_{o}$ at the peak emission wavelength. 
At large distances, the strength of the dipolar near-field of the molecule decreases in an inverse power law fashion, causing both the Purcell factor and the dissipation rate to decrease accordingly. As a result, the antenna effect weakens, so that $Y\longrightarrow Y_{o}$.
Likewise, in Fig. \ref{f5} at $d>15$ nm, $Y$ approaches the same value regardless of the core-offset.   
For the core-offsets and MNP size we studied, the optimal range of CV molecule-nanoegg distance for quantum yield enhancement is $\sim 3-15$ nm. 
\begin{figure}[h!] 
	\centering
	\includegraphics[width = 0.7\linewidth]{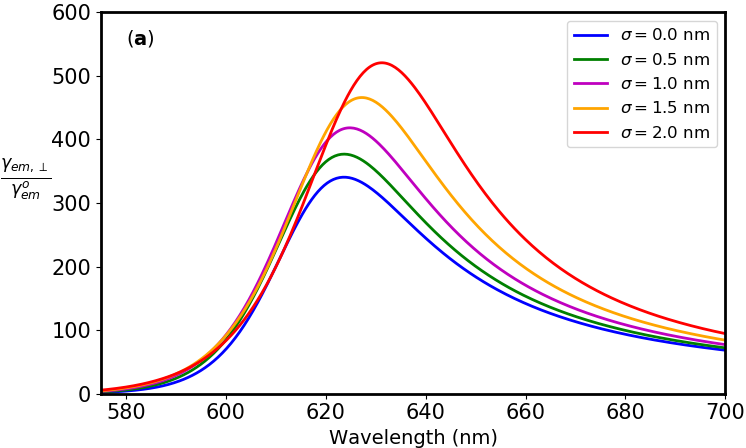}\vspace{0.2cm}\\
	\includegraphics[width = 0.7\linewidth]{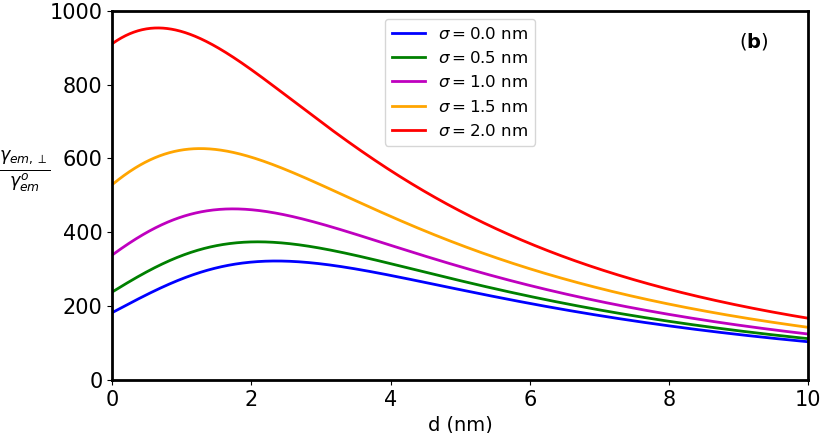}
	\caption{
		\small{Emission rate enhancement of a perpendicularly-oriented CV molecule for the core-offsets studied. 
			(\textbf{a}) At $d = 5$ nm from the nanoegg as a function of wavelength assuming that the excitation and emission wavelengths are
			 the same. (\textbf{b}) At the peak emission wavelength $\lambda = 640$ nm as a function of its distance from the nanoegg.}
	}
	\label{f6}
\end{figure}

We can predict the dependence of the emission rate of the molecule on the core-offset of the nanoegg by using a method
proposed in Ref. \cite{BN07}. It makes use of the optical reciprocity theorem, which gives that the Purcell factor 
and the excitation rate enhancement are identical for the perpendicular dipole \cite{BN07,Moha08}. Thus, if the molecule is excited at its peak emission wavelength \cite{BN07}, the Purcell factor is the same as the excitation rate enhancement. The result of this approach is shown in Fig. \ref{f6}, using Eq. (\ref{e95}). Fig. \ref{f6}~ shows that with increasing core-offset, the increase in excitation rate enhancement dominates the decrease in quantum yield enhancement. Hence, the emission spectrum of the molecule shows an increase in emission rate with increasing core-offset. 

Khatua et al. \cite{Khat14} reported an emission rate enhancement of $\sim 1000$ for a CV molecule at a distance of $5$ nm
from the tip of a gold nanorod. In Fig. \ref{f6}(\textbf{a}), our theoretical approach shows that at this distance, 
$\sim 50$\% of this enhancement factor can be achieved in a CV molecule-DCMS nanoegg system, via small core-offsets in a nanoshell with a radius comparable to the equivalent sphere-volume radius of the nanorod. However, Fig. \ref{f6}(\textbf{b}) shows that 
an emission rate enhancement of $\sim 900$ can be achieved in the CV molecule-DCMS nanoegg system at shorter distances 
via a core-offset of $\sigma = 2$ nm. At short distances $(d \longrightarrow 0)$, the emission rate enhancement does not decrease 
to zero because the excitation rate enhancement is maximum at $d = 0$ and $Y$ does not decrease to zero (Fig. \ref{f5}),
while at large distances, the emission rate enhancement tends to zero because the excitation rate enhancement approaches zero 
as $Y\longrightarrow Y_{o}$. A similar behaviour has been reported in the nanorod-CV molecule system of Ref. \cite{Khat14}. 

The dependence of the dielectric function of a MNP on the longitudinal propagation wavevector of the incident electric field 
causes gold nanoparticles to exhibit a size-dependent response, which differs from the bulk response given by Eq. (\ref{e1}). 
This non-local hehaviour places an upper bound on enhancement factors \cite{Cira12}. However, the trends in enhancement factors 
predicted by both the local and the non-local response remain the same. A major difference exists only in the lower values 
of enhancement factors and optimal MNP-molecule distances, as well as size-dependent spectral shifts, predicted by the non-local response \cite{Chris16,Cira12}. 

\section{Conclusion}\label{s4}
Off-setting the core of a DCMS nanoshell, embedded in a dielectric medium and placed near an excited molecule, 
causes both the dipolar near-field of the molecule and the dipolar surface plasmon mode to couple to all surface plasmon modes of the nanoshell. This process leads to the formation of dipole-active modes in the nanoshell, which increases the induced dipole moment on the molecule. As a result, the Purcell factor of the molecule increases with increasing core-offset. Likewise, the non-radiative energy transfer rate from the molecule to the nanoegg also increases, reaching a maximum at the dipolar LSPR of the nanoegg. 

Within the quasistatic limit, we have investigated the impact of these dipole-active modes on the radiative decay rate and quantum yield of a CV molecule placed near the nanoegg. The theoretical model we adopted shows that the nanoegg is a more efficient antenna for enhancement of the radiative decay rate compared to the concentric nanoshell, while the concentric nanoshell is a more efficient antenna for enhancement of quantum yield. However, a method based on optical reciprocity, has shown that the emission rate of the molecule is more enhanced near the nanoegg due to the dominant contribution from the excitation rate enhancement. 

We have considered both the normal and tangential orientations of the dipole moment of the CV molecule with respect to the nanoegg surface. 
We found that the tangential dipole is less enhanced. In addition, the peak wavelengths of the optimal enhancement factors are redshifted from the dipolar LSPR of the nanoegg. This result is consistent with PEF calculations in other plasmonic nanostructures within the LRA. 

Compared to a nanorod-emitter system, our theoretical model also shows that by using a nanoegg whose size is comparable to the nanorod, 
it is possible to achieve similar fluorescence enhancement factors, via large core-offsets in a nanoegg-emitter system.

\small{
\begin{acknowledgments}
L. C. U. was sponsored by the National Research Foundation (NRF) and the University of Pretoria.
T. M. was supported by the Czech Science Foundation (GACR) grant no. 18-18022S. 
T. P. J. K. was supported by the NRF grant nos. N01564 (project 109302) and N00500 (project 112085).
\end{acknowledgments}
}

\bibliography{Manus}
\end{document}